\documentclass[aps,superscriptaddress]{revtex4-2}
\usepackage[colorlinks=true, pdfstartview=FitV, linkcolor=blue, citecolor=red, urlcolor=magenta, breaklinks=true]{hyperref}
\usepackage{graphicx}  
\usepackage{amsfonts}
\begin{document}
\title{Modified metrics of acoustic black holes: A review}
\author{M. A. Anacleto}
\email{anacleto@df.ufcg.edu.br}
\affiliation{Departamento de F\'{\i}sica, Universidade Federal de Campina Grande
Caixa Postal 10071, 58429-900 Campina Grande, Para\'{\i}ba, Brazil}
\author{F. A. Brito}
\email{fabrito@df.ufcg.edu.br}
\affiliation{Departamento de F\'{\i}sica, Universidade Federal de Campina Grande
Caixa Postal 10071, 58429-900 Campina Grande, Para\'{\i}ba, Brazil}
\affiliation{Departamento de F\'isica, Universidade Federal da Para\'iba, 
Caixa Postal 5008, 58051-970 Jo\~ao Pessoa, Para\'iba, Brazil}
\author{E. Passos}
\email{passos@df.ufcg.edu.br}
\affiliation{Departamento de F\'{\i}sica, Universidade Federal de Campina Grande
Caixa Postal 10071, 58429-900 Campina Grande, Para\'{\i}ba, Brazil}

\begin{abstract} 

In this brief review, we will address acoustic black holes arising from quantum field theory in the Lorentz-violating 
and non-commutative background. 
Thus, we consider canonical acoustic black holes with effective metrics for the purpose of investigating Hawking radiation and entropy.
We show that due to the generalized uncertainty principle and the modified dispersion relation, the Hawking temperature is regularized, that is, free from the singularity when the horizon radius goes to zero. 
In addition, we also find logarithmic corrections in the leading order for entropy.

\end{abstract}
\maketitle
\pretolerance10000

\section{Introduction}
Gravitational analogue models are topics of great interest and have been widely studied in the literature due to the possibility of detecting Hawking radiation in the table experiment. 
In particular, acoustic black holes were proposed by Unruh in 1981~\cite{Unruh:1980cg,Unruh:1994je} for the purpose of exploring Hawking radiation, as well as investigating other issues to understand quantum gravity effects. 
It is well known that an acoustic black hole can be generated when fluid motion reaches a speed greater than the local speed of sound. 
These objects can exhibit properties similar to the laws of thermodynamics of gravitational black holes, such as a Hawking-like temperature and entropy (entanglement entropy). 
Besides, it has been conjectured that phenomena that are observed in black holes may also occur in acoustic black holes. 
Furthermore, with the detection of gravitational waves~\cite{LIGOScientific:2016aoc,LIGOScientific:2017vwq} and the capture of the image of a supermassive black hole~\cite{event2019firstI,event2019firstVI}, a window of possibilities in the physics of black holes and also in analogous models was opened. 
Acoustic black holes have applications in various branches of physics, namely high energy physics, 
condensed matter, and quantum physics~\cite{Visser:1997ux,Barcelo:2005fc}. 
On the experimental side, Hawking radiation has been successfully measured in the works reported 
in~\cite{MunozdeNova:2018fxv,Isoard:2019buh}. 
And also carried out in other branches of physics~\cite{Steinhauer:2014dra,Drori:2018ivu,Rosenberg:2020jde,Guo:2019tmr,Bera:2020doh,Blencowe:2020ygo}. 
However, in the physics of acoustic black holes, the first experimental measurement of Hawking radiation was devised in the Bose-Einstein condensate~\cite{Lahav:2009wx}. 

In a recent paper, acoustic black holes embedded in a curved background were constructed by applying relativistic Gross-Pitaevskii and Yang-Mills theories~\cite{Ge:2019our}. 
In~\cite{Yu:2017bnu}, an acoustic black hole of a D3-black brane was proposed. 
On the other hand, relativistic acoustic black holes in Minkowski spacetime were generated 
from the Abelian Higgs model~\cite{Ge:2010wx,Anacleto:2010cr,Anacleto:2011bv,Anacleto:2013esa,Anacleto:2021nhm}. 
Also, relativistic acoustic black holes have emerged from other physical models~\cite{Bilic:1999sq,Fagnocchi:2010sn,Giacomelli:2017eze,Visser:2010xv}. 
In addition, these objects have been used to analyze various physical phenomena, such as superradiance~\cite{Basak:2002aw,Richartz:2009mi,Anacleto:2011tr,Zhang:2011zzh,Ge:2010eu}, entropy~\cite{Zhao:2012zz,Anacleto:2014apa,Anacleto:2015awa,Anacleto:2016qll,Anacleto:2019rfn}, quasinormal modes~\cite{Cardoso:2004fi,Nakano:2004ha,Berti:2004ju,Chen:2006zy,Guo:2020blq,Ling:2021vgk}, and as well as, in other models~\cite{Dolan:2011zza,Anacleto:2012ba,Anacleto:2012du,Anacleto:2015mta,Anacleto:2016ukc,Anacleto:2018acl,Anacleto:2020kxj,Anacleto:2021wmv,Qiao:2021trw,Vieira:2014rva,Ribeiro:2021fpk}. 
Moreover, in~\cite{Zhang:2016pqx}, was reported that there is a thermodynamic-like description for acoustic black holes in two dimensions. In this sense, an analogous form of Bekenstein-Hawking entropy (understood as an entanglement entropy) was addressed in~\cite{Rinaldi:2011nb} by analyzing the Bose-Einstein condensate system. 
In addition, the dependence of entropy on the area of the event horizon of the acoustic black hole was explored in~\cite{Steinhauer:2015ava}. 
Also, in~\cite{Giovanazzi:2011az}, the entanglement entropy of an acoustic black hole was examined. 

In this brief review, we are interested in investigating modified acoustic black holes that have been constructed from field theory by considering the Abelian Higgs model in the Lorentz-violating~\cite{Anacleto:2010cr} 
and noncommutative~\cite{Anacleto:2011bv} background. 
To this end, we will explore canonical acoustic black holes with modified metrics to examine the effect of Lorentz symmetry breaking and noncommutativity on Hawking radiation and entropy.
In addition, by applying the generalized uncertainty principle and a modified dispersion relation, we show that the Hawking temperature singularity disappears when the horizon radius vanishes. 
Besides, we also find logarithmic correction terms for entropy. 
Recently, the stability of the canonical acoustic black hole in the presence of noncommutative effects and minimum length has been addressed by us in~\cite{Anacleto:2022lnt}.
Thus, it was verified that the non-commutativity and the minimum length act as regulators in the Hawking temperature, that is, the singularity is removed. 
Also, it was shown that for a certain minimum radius the canonical acoustic black hole presents stability.

This brief review is organized as follows. In Sec.~\ref{ABH}, we  briefly review the steps to find the relativistic acoustic black hole metrics. 
In Sec.~\ref{MABH}, we  briefly review the steps to find the relativistic acoustic black hole modified metrics. 
In Sec.~\ref{MCABH}, wwe will focus on canonical acoustic black holes with effective metrics to compute Hawking temperature and entropy. 
In Sec.~\ref{QCTHE}, we will introduce quantum corrections via the generalized uncertainty principle and the modified dispersion relation in the calculation of Hawking temperature and entropy.
Finally in Sec.~\ref{conclu} we present our final considerations.

\section{Acoustic Black Hole}
\label{ABH}
In this section we review the steps to obtain the relativistic acoustic metric from the Lagrangian density of the charged scalar field. Here we will follow the procedure adopted in~\cite{Ge:2010wx}.
\subsection{Relativistic Acoustic Metric}
{In order to determine the relativistic acoustic metric, we start by considering the following Lagrangian density:
\begin{eqnarray}
\label{lsf}
{\cal L}=\partial_{\mu}\phi^{\ast}\partial^{\mu}\phi+ m^2|\phi|^2-b|\phi|^4.
\end{eqnarray}
Now, we decompose the scalar field as  $\phi = \sqrt{\rho(x, t)} \exp {(iS(x, t))}$, such that
\begin{eqnarray}
&&{\cal L} = \rho\partial_{\mu}S\partial^{\mu}S + m^2\rho - b\rho^2
+\frac{\rho}{\sqrt{\rho}}(\partial_{\mu}\partial^{\mu})\sqrt{\rho}.
\end{eqnarray}
Moreover, from the above Lagrangian, we find the equations of motion for $ S $ and $ \rho $ given respectively by
\begin{eqnarray}
\label{cont0}
\partial_{\mu}\left(\rho\partial^{\mu}{S}\right)=0,
\end{eqnarray}
and
\begin{eqnarray}
\label{fluid0}
\frac{1}{\sqrt{\rho}}{\partial_{\mu}\partial^{\mu}\sqrt{\rho}}
+\partial_{\mu}S\partial^{\mu}{S}+m^2-2b\rho=0,
\end{eqnarray}
where the Eq. (\ref{cont0}) is the continuity equation and Eq. (\ref{fluid0}) is an equation describing a hydrodynamical fluid, and the term, $ \frac{1}{\sqrt{\rho}}{\partial_{\mu}\partial^{\mu}\sqrt{\rho}} $, called the quantum potential can be neglected in the hydrodynamic region.

Now, by performing the following perturbations on equations of motion (\ref{cont0}) and (\ref{fluid0}):
\begin{eqnarray}
\rho=\rho_0+ \epsilon\rho_1 + \mathcal{O}(\epsilon^2),
\\
S=S_0+\epsilon\psi + \mathcal{O}(\epsilon^2).
\end{eqnarray}
We obtain
\begin{eqnarray}
\partial _\mu \left( \rho _1 u_0^{\mu}
+\rho _0\partial ^\mu \psi\right) =0,
\label{13}
\end{eqnarray}
and
\begin{eqnarray}
u_0^{\mu} \partial _\mu\psi -b\rho _1 =0,
\label{15}
\end{eqnarray}
where we have defined $ u_0^\mu =\partial ^\mu S_0$.
Hence, solving (\ref{15}) for $\rho_1$ and substituting into (\ref{13}), we have
\begin{eqnarray}
\partial _\mu 
\left[u^{\mu}_0u^{\nu}_0 + b\rho _0 g^{\mu\nu}\right]\partial_\nu\psi  =0.
\end{eqnarray}

We can also write the above equation as follows:
\begin{eqnarray}
\label{eqwave}
&&\partial _t\left\lbrace  {\omega}^2_0 
\left[ -1- \frac{b\rho_0}{2{\omega}^2_0}\right]\partial _t \psi 
-{\omega}^2_0\frac{v^i_0}{{\omega}_0}\partial _i \psi \right\rbrace 
\nonumber\\
&& +\partial _i\left\lbrace {-\omega}^2_0 \frac{v^i_0}{{\omega}_0}\partial _t \psi
%\nonumber\\
+{\omega}^2_0\left[ -\frac{ v^i_0 v^j_0}{{\omega}^2_0} +\frac{b\rho_0}{2{\omega}^2_0} \delta^{ij}\right]
\partial _j \psi \right\rbrace  =0,
\end{eqnarray}
where $ \omega_0=-\partial^t S_0 $ and $v_0^i=\partial _i S_0$ (the local velocity field).
In addition, we define 
$ c^2_s=b\rho_0/2\omega^2_0 $ to be the speed of sound
and $ v^i=v^i_0/\omega_0 $.
However, the equation (\ref{eqwave}) becomes
\begin{eqnarray}
\label{KG}
&&\partial _t\left\lbrace \frac{b\rho _0}{2c_s ^2}
\left[ \left(-1 -c_s^2 \right)\partial _t \psi 
- v^i \partial _i\psi\right] \right\rbrace 
%\nonumber\\
%&& 
+\, \partial _i\left\lbrace \frac{b\rho _0}{2c_s^2}
\left[ - v^i\partial _t \psi 
+\left( - v^i v^j +c_s^2 \delta^{ij}\right)
\partial _j \psi \right]\right\rbrace =0.
\end{eqnarray}
In this way, the above equation can be written as a Klein-Gordon equation in (3+1) dimensional curved space as follows:
\begin{eqnarray}
\frac{1}{\sqrt{-g}}\partial_{\mu}\left( \sqrt{-g}g^{\mu\nu}\partial_{\nu}\right)\psi=0,
\end{eqnarray}
where
\begin{eqnarray}
\sqrt{-g}g^{\mu\nu}=\frac{b\rho_0}{2c^2_s}
\left(\begin{array}{ccc}
-1-c^2_s  & \vdots & -v^{i} \\ 
\cdots\cdots &\cdot &\cdots\cdots\\
-v^{j} & \vdots & c^2_s\delta^{ij} - v^i v^j
\end{array} 
\right).
\end{eqnarray}
Hence, by determining the inverse of $ g^{\mu\nu} $, we find the relativistic acoustic metric given by
\begin{eqnarray}
\label{relmt}
g_{\mu\nu}=\frac{b\rho_0}{2c_s\sqrt{1+c^2_s-v^2}} 
\left(\begin{array}{ccc}
-c^2_s + v^2 & \vdots & -v^{i} \\ 
\cdots\cdots &\cdot &\cdots\cdots\\
-v^{j} & \vdots & (1+c^2_s)\delta^{ij}
\end{array} 
\right).
\end{eqnarray}
The metric depends on the density $\rho_0$, the local sound speed in the fluid $c_s$, the velocity of flow $\vec{v}$.
This is the acoustic black hole metric for high $c_s$ and $\vec{v}$ speeds.
Note that, in the non-relativistic limit, up to a overall factor, the metric found by Unruh is obtained.
\begin{eqnarray}
g_{\mu\nu}=\frac{b\rho_0}{2c_s} 
\left(\begin{array}{ccc}
-c^2_s + v^2 & \vdots & -v^{i} \\ 
\cdots\cdots &\cdot &\cdots\cdots\\
-v^{j} & \vdots & \delta^{ij}
\end{array} 
\right).
\end{eqnarray}
The relativistic acoustic metric (\ref{relmt}) has also been obtained from the Abelian Higgs model~\cite{Ge:2010wx}.

\subsection{The Dispersion Relation}
\label{dispersion}
Here we aim to examine the dispersion relation. 
Hence, we will adopt the notation written below
\begin{eqnarray}
\psi\sim \mbox{Re}\left[e^{i{\omega} t - i \vec{k}\cdot\vec{x}}\right],
\qquad
{\omega}=\frac{\partial\psi}{\partial t},
\qquad
\vec{k}=\nabla\psi.
\end{eqnarray}
So we can write the Klein-Gordon equation (\ref{KG}) in terms of momentum and frequency as follows:
\begin{eqnarray}
(1+c^2_s)\,{\omega}^2 + 2(\vec{v}\cdot\vec{k})\,{\omega} 
-\left( c_s^2 - v^2\right) k^2 = 0.
\end{eqnarray}
Now, by making $ k^i=\delta^{i1} $, we have
\begin{eqnarray}
{\omega}=\frac{-v_1 k\,\pm c_s k\sqrt{1+c^2_s - v^2_1}}{(1+c^2_s)}
=\frac{-v_1 k\,\pm c_s k\sqrt{1+(c_s - v_1)(c_s + v_1)}}{(1+c^2_s)},
\end{eqnarray}
In the limit of small $ v_1 $, we find the modified dispersion relation
\begin{eqnarray}
\label{rdisp}
\omega\approx E\left(1+\frac{v_1}{2} \right),
\end{eqnarray}
where $ E=c_s\,k $ is the linear dispersion relation. }

\section{Modified Acoustic Black Hole}
\label{MABH}
In this section we review the derivation of the relativistic acoustic metric from the Abelian Higgs model in the background violating-Lorentz and noncommutative.
\subsection{The Lorentz-Violating Model}
\label{IIa}
At this point, we consider the Abelian Higgs model with Lorentz symmetry breaking that has been introduced as a change in the scalar sector of the Lagrangian~\cite{Bazeia:2005tb}.
Moreover, the relativistic acoustic metric violating Lorentz has been found in~\cite{Anacleto:2010cr}.
Then, the corresponding Lagrangian for the abelian Higgs model in the Lorentz-violating background is written as follows:
\begin{eqnarray}
\label{acao}
{\cal L}&=&-\frac{1}{4}F_{\mu\nu}F^{\mu\nu} +|D_{\mu}\phi|^2+ m^2|\phi|^2-b|\phi|^4+ k^{\mu\nu}D_{\mu}\phi^{\ast}D_{\nu}\phi,
\end{eqnarray}
being $F_{\mu\nu}=\partial_{\mu}A_{\nu}-\partial_{\nu}A_{\mu}$ the field intensity tensor, $D_{\mu}\phi=\partial_{\mu}\phi - ieA_{\mu}\phi$ the covariant derivative and $k^{\mu\nu}$  a constant tensor implementing the Lorentz symmetry breaking, 
given by~\cite{Anacleto:2010cr}
\begin{equation}
k_{\mu\nu}=\left[\begin{array}{clcl}
\beta &\alpha &\alpha & \alpha\\
\alpha &\beta &\alpha &\alpha \\
\alpha &\alpha &\beta &\alpha\\
\alpha &\alpha &\alpha &\beta
\end{array}\right], \quad(\mu,\nu=0,1,2,3),
\end{equation}
where $\alpha$ and $\beta$ are real parameters. 

Next, following the steps taken in the previous section to derive the relativistic acoustic metric from quantum field theory, we consider $\phi = \sqrt{\rho(x, t)} \exp {(iS(x, t))}$ in the Lagrangian above. Thus, we have
\begin{eqnarray}
{\cal L} &=& -\frac14 F_{\mu\nu}F^{\mu\nu} + \rho\partial_{\mu}S\partial^{\mu}S - 2e\rho A_\mu \partial^{\mu}S+ e^2\rho A_\mu A^\mu + m^2\rho - b\rho^2\nonumber\\
&+& k^{\mu\nu} \rho(\partial_{\mu}S\partial_{\nu}S-2eA_\mu\partial_{\nu}S+e^2 A_\mu A_\nu)
+\frac{\rho}{\sqrt{\rho}}(\partial_{\mu}\tilde{\partial}^{\mu})\sqrt{\rho},
\end{eqnarray}
where $ \tilde{\partial}^{\mu}=\partial^\mu + k^{\mu\nu}\partial_\nu $. The equations of motion for $ S $ and $ \rho $ are:
\begin{eqnarray}
\label{cont}
\partial_{\mu}\left[\rho u^{\mu} 
+ \rho k^{\mu\nu}u_{\nu}\right]=0,
\end{eqnarray}
and
\begin{eqnarray}
\label{fluid}
\frac{(\partial_{\mu}\tilde{\partial}^{\mu})\sqrt{\rho}}{\sqrt{\rho}}
+ u_{\mu}u^{\mu} + k^{\mu\nu}u_{\mu}u_{\nu} +m^2 - 2b\rho=0,
\end{eqnarray}
where we have defined $ u^{\mu} =\partial^{\mu} S - e A^{\mu}$.
Now, by linearizing the equations above around the background $(\rho_0,S_0)$, with 
\begin{eqnarray}
\rho=\rho_0+ \epsilon\rho_1 + \mathcal{O}(\epsilon^2),
\\
S=S_0+\epsilon\psi + \mathcal{O}(\epsilon^2),
\end{eqnarray} 
and keeping the vector field $A_{\mu}$ unchanged, we have
\begin{eqnarray}
\label{eqcvla}
\partial_{\mu} \left[\rho_1\left( u_0^{\mu}+ k^{\mu\nu}u_{0\nu} \right)
+\rho_0\left(g^{\mu\nu}+k^{\mu\nu}\right)\partial_{\nu}\psi\right] = 0,
\end{eqnarray}
and
\begin{eqnarray}
\label{eqhvlb}
\left( u_0^{\mu} + k^{\mu\nu}u_{0\nu}\right) \partial _\mu\psi -b\rho _1 =0,
\end{eqnarray}
by solving (\ref{eqhvlb}) for $ \rho_1 $ and replacing into equation (\ref{eqcvla}), we obtain
\begin{eqnarray}
\partial _\mu 
\left[u^{\mu}_0u^{\nu}_0 + k^{\mu\lambda}u_{0\lambda}u_{0}^{\nu} + u_{0}^{\mu}k^{\nu\lambda}u_{0\lambda} 
+ b\rho _0 \left(g^{\mu\nu}+k^{\mu\nu}\right)\right]\partial_\nu\psi  =0.
\end{eqnarray}
Hence, we find the equation of motion for a linear acoustic disturbance $\psi$ given by a Klein-Gordon equation in a curved space
\begin{eqnarray}
\frac{1}{\sqrt{-g}}\partial_{\mu}(\sqrt{-g}g^{\mu\nu}\partial_{\nu})\psi=0,
\end{eqnarray}
where $g_{\mu\nu}$ is the relativistic acoustic metrics. 

For $\beta\neq0$ and $\alpha=0$, we have~\cite{Anacleto:2010cr}
\begin{equation}
\label{invs_g}
g_{\mu\nu}\equiv\frac{b\rho_{0}\tilde{\beta}_{-}^{1/2}}{2c_{s}\sqrt{\mathcal{Q}}}
\left[\begin{array}{clcl}
-\left(\frac{c_{s}^2}{\tilde{\beta}_{+}}-\frac{\tilde{\beta}_{-}}{\tilde{\beta}_{+}}v^2\right)& \vdots &\quad\quad\quad\quad-v^{j}\\
\cdots\cdots\cdots\cdots\cdots &\cdot & \quad\quad\cdots\cdots\cdots\cdots\cdots\cdots\\
-v^{i} & \vdots & f_{\beta}\delta^{ij}+\frac{\tilde{\beta}_{-}}{\tilde{\beta}_{+}}v^{i}v^{j}
\end{array}\right],
\end{equation}
where ${\cal Q}=1+\frac{c_{s}^2}{\tilde{\beta}_{+}}-\frac{\tilde{\beta}_{-}}{\tilde{\beta}_{+}}v^2$
and $f_{\beta}=\frac{\tilde{\beta}_{+}}{\tilde{\beta}_{-}}
+\frac{c_{s}^2}{\tilde{\beta}_{-}}-\frac{\tilde{\beta}_{-}}{\tilde{\beta}_{+}}v^2$.  

The acoustic line element in the Lorentz-violating background can be written as follows
\begin{eqnarray}
ds^2&=&\frac{b\rho_{0}\tilde{\beta}_{-}^{1/2}}{2c_{s}\sqrt{\cal{Q}}}
\left[-\left(\frac{c_{s}^2}{\tilde{\beta}_{+}}-\frac{\tilde{\beta}_{-}}{\tilde{\beta}_{+}}v^2\right)dt^2-2\vec{v}\cdot d\vec{x}dt
+\frac{\tilde{\beta}_{-}}{\tilde{\beta}_{+}}(\vec{v}\cdot d\vec{x})^2
+f_{\beta} d\vec{x}^2\right].
\end{eqnarray}
Now changing the time coordinate as $d\tau=dt + \frac{\tilde{\beta}_{+}\vec{v}\cdot d\vec{x}}{c^2_{s}-\tilde{\beta}_{-}v^2}$, 
we find the acoustic metric in the stationary form
\begin{eqnarray}
ds^2&=&\frac{b\rho_{0}\tilde{\beta}_{-}^{1/2}}{2c_{s}\sqrt{{\cal Q}}}
\left[-\left(\frac{c_{s}^2}{\tilde{\beta}_{+}}-\frac{\tilde{\beta}_{-}}{\tilde{\beta}_{+}}v^2\right)d\tau^2+
{\cal F}\left(\frac{\tilde{\beta}_{-}v^{i}v^{j}}{c^2_{s}-\tilde{\beta}_{-}v^2}
+\frac{f_{\beta}}{{\cal F}}\delta^{ij}\right)dx^{i}dx^{j}\!\right]\!\!.
\end{eqnarray}
where ${\cal F}=\left(\frac{\tilde{\beta}_{+}}{\tilde{\beta}_{-}}+\frac{c_{s}^2}{\tilde{\beta}_{+}}
-\frac{\tilde{\beta}_{-}}{\tilde{\beta}_{+}}v^2\right)$. 
For $\tilde{\beta}=1$ we recover the result found in Ref.~\cite{Ge:2010wx}.

Next, for $\beta=0$ and $\alpha\neq 0$, we have~\cite{Anacleto:2010cr}
\begin{equation}
\label{met}
g_{\mu\nu}\equiv\frac{b\rho_{0}}{2c_{s}\sqrt{f}}
\left[\begin{array}{clcl}
g_{tt} &\vdots &g_{tj}\\
\cdots&\cdot&\cdots\\
g_{it} &\vdots &g_{ij}
\end{array}\right],
\end{equation}
where
\begin{eqnarray}
\label{gtt}
g_{tt}&=&-[(1+\alpha)c^2_{s}-v^2+\alpha^2(1-v)^2],
\\
g_{tj}&=&-(1-\vec{\alpha}\cdot\vec{v})v^{j},
\\
g_{it}&=&-(1-\vec{\alpha}\cdot\vec{v})v^{i},
\\
g_{ij}&=&\left[(1-\vec{\alpha}\cdot\vec{v})^2+c^2_{s}-v^2\right]\delta^{ij} +v^{i}v^{j},
\label{gij}
\\
f&=&(1+\alpha)[(1-\vec{\alpha}\cdot\vec{v})^2+c^2_{s}]-v^2+\alpha^{2}(1-v)^2\left[1+(1-\vec{\alpha}\cdot\vec{v})^2c^{-2}_{s}\right].
\end{eqnarray}
Thus, the acoustic line element in the Lorentz-violating background can be written as
\begin{eqnarray}
ds^2&=&\frac{b\rho_{0}}{2c_{s}\sqrt{f}}
\left[g_{tt}dt^2-2(1-\vec{\alpha}\cdot\vec{v})(\vec{v}\cdot d\vec{x})dt+(\vec{v}\cdot d\vec{x})^2
+ f_{\alpha} d\vec{x}^2\right],
\end{eqnarray}
where $ f_{\alpha}=(1-\vec{\alpha}\cdot\vec{v})^2+c^2_{s}-v^2 $. Now changing the time coordinate as 
\begin{equation}
d\tau=dt + \frac{(1-\vec{\alpha}\cdot\vec{v})(\vec{v}\cdot d\vec{x})}
{[(1+\alpha)c^2_{s}-v^2+\alpha^2(1-v)^2]},
\end{equation} 
we find the acoustic metric in the stationary form
\begin{eqnarray}
ds^2=\frac{b\rho_{0}}{2c_{s}\sqrt{f}}
\left[g_{tt}d\tau^2+\Lambda\left(\frac{-v^{i}v^{j}}{g_{tt}}
+\frac{f_{\alpha}\delta^{ij}}
{\Lambda}\right)dx^{i}dx^{j}\right]\!.
\end{eqnarray}
where $ \Lambda=(1-\vec{\alpha}\cdot\vec{v})^2-g_{tt} $.
For $\alpha=0$, the result found in~\cite{Ge:2010wx} is recovered. 

\subsection{Noncommutative Acoustic Black Hole}
\label{II}
The metric of a noncommutative canonical acoustic black hole has been found by us in~\cite{Anacleto:2011bv}.
Here, starting from the noncommutative Abelian Higgs model, we briefly review the steps to generate the relativistic acoustic metric in the noncommutative background.
Thus, the Lagrangian of the Abelian Higgs model in the noncommutative background is given by~\cite{Ghosh:2004wi}
\begin{eqnarray}
\label{acao}
\hat{\cal L}&=&-\frac{\kappa_+}{4}F_{\mu\nu}F^{\mu\nu}
+\kappa_-\left(|D_{\mu}\phi|^2+ m^2|\phi|^2-b|\phi|^4\right)
%\nonumber\\
+\frac{1}{2}\theta^{\alpha\beta}F_{\alpha\mu}\left[(D_{\beta}\phi)^{\dagger}D^{\mu}\phi+(D^{\mu}\phi)^{\dagger}D_{\beta}\phi \right],
\end{eqnarray}
being $ \kappa_{\pm}=1 \pm \theta^{\mu\nu}F_{\mu\nu}/2$, $F_{\mu\nu}=\partial_{\mu}A_{\nu}-\partial_{\nu}A_{\mu}$ the field intensity tensor  
and $D_{\mu}\phi=\partial_{\mu}\phi - ieA_{\mu}\phi$ the covariant derivative. 
The parameter $\theta^{\alpha\beta}$ is a constant, real-valued antisymmetric $D\times D$- matrix in $D$-dimensional spacetime with dimensions of length squared. 

Now, we use $\phi = \sqrt{\rho(x, t)} \exp {(iS(x, t))}$ in the above Lagrangian, 
such that~\cite{Anacleto:2011bv}.
\begin{eqnarray}
\label{lagran}
{\cal L}&=&-\frac{\kappa_+}{4}F_{\mu\nu}F^{\mu\nu}
+\rho\bar{g}^{\mu\nu}{\cal D}_{\mu}S{\cal D}_{\nu}S+\tilde{\theta} m^2\rho-\tilde{\theta}b\rho^2
+\frac{\rho}{\sqrt{\rho}}\bar{g}^{\mu\nu}\partial_{\mu}\partial_{\nu}\sqrt{\rho},
\end{eqnarray}
where ${\cal D}_{\mu}=\partial_{\mu}-eA_{\mu}/S  $, $ \bar{g}^{\mu\nu}=\tilde{\theta}g^{\mu\nu}+\Theta^{\mu\nu} $, $\tilde{\theta}=(1+\vec{\theta}\cdot\vec{B})$, $\vec{B}=\nabla\times\vec{A}$ and $\Theta^{\mu\nu}=\theta^{\alpha\mu}{F_{\alpha}}^{\nu}$. 
In our analysis we consider the case where there is no noncommutativity between space and time, that is $\theta^{0i}=0$ and use $\theta^{ij}=\varepsilon^{ijk}\theta^{k}$, $F^{i0}=E^{i}$ and $F^{ij}=\varepsilon^{ijk}B^{k}$.

In the sequence we obtain the equations of motion for $ S $ and $\rho$ as follows:
\begin{eqnarray}
&&\partial_{\mu}\left[\tilde{\theta}\rho u^{\mu}
+\rho\tilde{\Theta}^{\mu\nu}u_{\nu}\right]=0,
\end{eqnarray}
and
\begin{eqnarray}
&&\frac{1}{\sqrt{\rho}}\bar{g}^{\mu\nu}\partial_{\mu}{\partial}_{\nu}\sqrt{\rho}
+\bar{g}^{\mu\nu}u_{\mu}u_{\nu}
+\tilde{\theta}m^2-2\tilde{\theta}b\rho=0,
\end{eqnarray}
where $ \tilde{\Theta}^{\mu\nu}=(\Theta^{\mu\nu}+\Theta^{\nu\mu})/2 $.
Hence, by linearizing the equations of motion around the background $(\rho_0,S_0)$, with $\rho=\rho_0+\rho_1$,  $S=S_0+\psi$ 
and keeping the vector potential $ A_{\mu} $ unchanged, such that
\begin{eqnarray}
\label{eqcnca}
\partial_{\mu} \left[\rho_1\bar{g}^{\mu\nu}u_{0\nu}
+\rho_0\left(g^{\mu\nu}+\tilde{\Theta}^{\mu\nu}\right)\partial_{\nu}\psi\right] = 0,
\end{eqnarray}
and
\begin{eqnarray}
\label{eqhncb}
\left(\tilde{\theta} u_0^{\mu} + \tilde{\Theta}^{\mu\nu}u_{0\nu}\right) \partial _\mu\psi -b\tilde{\theta}\rho _1 =0.
\end{eqnarray}
Then, by manipulating the above equations, we obtain the equation of motion for a linear acoustic disturbance $\psi$ in the form 
\begin{eqnarray}
\frac{1}{\sqrt{-g}}\partial_{\mu}(\sqrt{-g}g^{\mu\nu}\partial_{\nu})\psi=0,
\end{eqnarray}
where $g_{\mu\nu}=\frac{b\rho_0}{2c_s\sqrt{f}}\tilde{g}_{\mu\nu}$ is the relativistic acoustic metric with noncommutative corrections in (3+1) dimensions and with $ \tilde{g}_{\mu\nu} $ given in the form~\cite{Anacleto:2011bv}
\begin{eqnarray}
\tilde{g}_{tt}&=&-[(1-3\vec{\theta}\cdot\vec{B})c^{2}_{s}-(1+3\vec{\theta}\cdot\vec{B})v^2
+2(\vec{\theta}\cdot\vec{v})(\vec{B}\cdot\vec{v})-(\vec{\theta}\times\vec{E})\cdot\vec{v}],
%\nonumber
\\
\tilde{g}_{tj}&=&-\frac{1}{2}(\vec{\theta}\times\vec{E})^{j}(c^{2}_{s}+1)-\left[2(1+2\vec{\theta}\cdot\vec{B})
-(\vec{\theta}\times\vec{E})\cdot\vec{v}\right]\frac{v^j}{2}+\frac{B^j}{2}(\vec{\theta}\cdot\vec{v})+\frac{\theta^j}{2}(\vec{B}\cdot\vec{v}),
%\nonumber
\\
\tilde{g}_{it}&=&-\frac{1}{2}(\vec{\theta}\times\vec{E})^{i}(c^{2}_{s}+1)-\left[2(1+2\vec{\theta}\cdot\vec{B})-(\vec{\theta}\times\vec{E})\cdot\vec{v}\right]\frac{v^i}{2}
+\frac{B^i}{2}(\vec{\theta}\cdot\vec{v})+\frac{\theta^i}{2}(\vec{B}\cdot\vec{v}),
%\nonumber
\\
\tilde{g}_{ij}&=&[(1+\vec{\theta}\cdot\vec{B})(1+c^2_{s})-(1+\vec{\theta}\cdot\vec{B})v^2
-(\vec{\theta}\times\vec{E})\cdot\vec{v}]\delta^{ij}+(1+\vec{\theta}\cdot\vec{B})v^{i}v^{j}.
%\nonumber
\\
f&=&[(1-2\vec{\theta}\cdot\vec{B})(1+c^2_{s})-(1+4\vec{\theta}\cdot\vec{B})v^2]
-3(\vec{\theta}\times\vec{E})\cdot\vec{v}+2(\vec{B}\cdot\vec{v})(\vec{\theta}\cdot\vec{v}).
\end{eqnarray}
Setting $ \theta=0 $, the acoustic metric above reduces to the acoustic metric obtained in Ref.~\cite{Ge:2010wx}.

\section{Modified canonical acoustic black hole}
\label{MCABH}
In this section, we shall address the issue of Hawking temperature in the regime of low velocities 
for the previous cases with further details. Now we consider an incompressible fluid with spherical symmetry. In this case the density $\rho$ is a position independent quantity and the continuity equation implies that $v\sim \frac{1}{r^2}$. The sound speed is also a constant. 
In the following we examine the Hawking radiation and entropy of the usual canonical acoustic black hole, as well as, in the Lorentz-violating and noncommutative background.

\subsection{Canonical Acoustic Metric}
In this case the line element of the acoustic black hole is given by
\begin{eqnarray}
ds^2=-f(v_{r})d\tau^2+\frac{c^2_{s}}{f(v_{r})}dr^2
+r^2(d\theta^2+\sin^2\theta d\phi^2),
\end{eqnarray}
where the metric function, $ f(v_{r}) $ takes the form
\begin{equation}
\label{hz1}
f(v_{r})={c^2-v^2_{r}}\quad\longrightarrow\quad f(r)=c^2_{s}\left(1-\frac{r^4_{h}}{r^4}\right).
\end{equation}
Here we have defined
$v_{r}=c_{s}\frac{r^2_{h}}{r^2}$,
being $r_{h}$ the radius of the event horizon. 

In this case we compute the Hawking temperature using the following formula:
\begin{equation}
\label{temph}
T_{H}=\frac{f^{\prime}(r_{h})}{4\pi}=\frac{c^2_{s}}{\pi r_{h}}.
\end{equation}
By considering the above result for the Hawking temperature and applying the first law of thermodynamics, we can obtain the entropy (entanglement entropy~\cite{Anacleto:2019rfn}) of the acoustic black hole as follows
\begin{eqnarray}
S=\int \frac{dE}{T}=\int \frac{dA}{4\pi r_h T_H}=\frac{A}{4c^2_s},
\end{eqnarray}
being $ A=4\pi r^2_h $ the horizon area of the canonical acoustic black hole.

\subsection{Canonical Acoustic Metric with Lorentz Violation}
%\noindent
In the limit $c^2_{s}\ll1$ and $v^2\ll1$ can be written as a Schwarzschild metric type. 
Thus, for $\beta\neq0$ and $\alpha=0$ and up to an irrelevant position-independent factor, 
we have~\cite{Anacleto:2010cr}
\begin{eqnarray}
ds^2&=&-f(v_{r})d\tau^2+\frac{c^2_{s}}{\sqrt{\tilde{\beta}_{-}\tilde{\beta}_{+}}f(v_{r})}dr^2
+\sqrt{\frac{\tilde{\beta_{+}}}{\tilde{\beta}_{-}}}r^2(d\theta^2+\sin^2\theta d\phi^2),
\end{eqnarray}
where
\begin{equation}
\label{hz1}
f(v_{r})=\sqrt{\frac{\tilde{\beta_{-}}}{\tilde{\beta}_{+}}}\left[\frac{c^2_{s}-\tilde{\beta}_{-}v^2_{r}}{\tilde{\beta}_{+}}\right]\rightarrow 
f(r)=\sqrt{\frac{\tilde{\beta_{-}}}{\tilde{\beta}_{+}}}
\left[\frac{c^2_{s}}{\tilde{\beta}_{+}}\left(1-\tilde{\beta}_{-}\frac{r^4_{h}}{r^4}\right)\right].
\end{equation}

The Hawking temperature is given by
\begin{equation}
T_{H}=\frac{f^{\prime}(r_{h})}{4\pi}=\frac{c^2_{s}(1-\beta)^{3/2}}{(1+\beta)^{3/2}\pi r_{h}}
=\frac{c^2_{s}(1-3\beta)}{\pi r_{h}}.
\end{equation}
Therefore, the temperature is decreased when we vary the parameter $\beta$.
For $\beta=0$ the usual result is obtained. 
Hence, from the above temperature, we have the following result for the entropy of the acoustic black hole in the background violating Lorentz.
\begin{eqnarray}
S=\frac{(1+3\beta)}{4c^2_s}A.
\end{eqnarray}

Now, for $\beta=0$ and $\alpha\neq 0$, we find for $\alpha$ sufficiently small we have up to first order
\begin{eqnarray}
&&f(v_{r})=\frac{\tilde{\alpha}c^2_{s}-v^2_r]}
{\sqrt{\tilde{\alpha}(1-2\alpha v_{r})}},
\end{eqnarray}
where $\tilde{\alpha}=1+\alpha$. For $v_r=c_s r^2_h/r^2$ with $ c_s=1 $, the metric function becomes
\begin{equation}
f(r)\simeq \tilde{\alpha}^{-1/2}\left[\tilde{\alpha}-\frac{r_h^4}{r^4}\left(1+\alpha \frac{r_h^2}{r^2} \right)
+\alpha\frac{r_h^2}{r^2}\right].
\end{equation}
In the present case there is a richer structure such as charged and rotating black holes. 
{The event horizon of the modified canonical acoustic black hole is obtained from the following equation:
\begin{eqnarray}
\tilde{\alpha}-\frac{r_h^4}{r^4}\left(1+\alpha \frac{r_h^2}{r^2} \right)
+\alpha\frac{r_h^2}{r^2}=0,
\end{eqnarray}
which can also be rewritten in the form
\begin{eqnarray}
r^6 + \alpha\, r^2_h\, r^4 - \tilde{\alpha}^{-1\,}r^4_ h\, r^2 - \alpha\, r^6_h=0, 
\end{eqnarray}
we can also write
\begin{eqnarray}
\label{eqq6}
r^2(r^2 - r^2_+)(r^2- r^2_{-}) - \alpha\, r^6_h=0,
\end{eqnarray}
where
\begin{eqnarray}
r^2_{\pm}=r^2_h\left(-\frac{\alpha}{2} \pm \frac{1}{\sqrt{\tilde{\alpha}}}\right).
\end{eqnarray}
Now, arranging the above equation (\ref{eqq6}), we have
\begin{eqnarray}
r^2=r^2_{+} + \frac{\alpha r^6_h}{r^2(r^2 - r^2_{-} )}.
\end{eqnarray} 
Therefore, we can find the event horizon by solving the above equation perturbatively. 
So, up to the first order in $\alpha$, we obtain
\begin{eqnarray}
\tilde{r}^2_{+}\approx r^2_{+} + \frac{\alpha r^6_h}{r^2_+(r^2_+ - r^2_{-} )}
=\left( 1-\frac{\alpha}{2} \right) r^2_h  + \cdots.
\end{eqnarray}
Then, we have
\begin{eqnarray}
\tilde{r}_+=r_h\sqrt{1-\frac{\alpha}{2}} + \cdots. 
\end{eqnarray}
   
For the Hawking temperature, we obtain
\begin{eqnarray}
\label{tha}
T_H=\frac{1}{\pi \tilde{r}_+}\left(1+\frac{3\alpha}{2}  \right).
\end{eqnarray}
In terms of $r_h$, we have}
\begin{equation}
T_{H}=\left(1+\frac{7\alpha}{4}\right)\frac{1}{(\pi r_h)}.
\end{equation}
In this situation the temperature is increased when we vary the parameter $\alpha$.
For $\alpha=0$ one recovers the usual result.

In this case for entropy, we find
\begin{eqnarray}
S=\left(1-\frac{7\alpha}{4}\right)\frac{A}{4}.
\end{eqnarray}

\subsection{Noncommutative Canonical Acoustic Metric}

The noncommutative acoustic metric can be written as a Schwarzschild metric type, up to an irrelevant
position-independent factor, in the nonrelativistic limit as follows~\cite{Anacleto:2011bv},
\begin{eqnarray}
ds^2&=&-\tilde{{\cal F}}(v_{r})d\tau^2+\frac{[v_{r}^2\Gamma+\Sigma+\tilde{{\cal F}}(v_{r})\Lambda]}{\tilde{\cal F}(v_{r})}dr^2
+\frac{r^2(d{\vartheta}^2+\sin^2\vartheta d\phi^2)}{\sqrt{f}},
\end{eqnarray}
where
\begin{eqnarray}
\tilde{{\cal F}}(v_{r})&=&\frac{{\cal F}(v_{r})}{\sqrt{f(v_r)}}=\frac{1}{\sqrt{f(v_r)}}
\left[(1-3\vec{\theta}\cdot\vec{B})c^2_{s}-(1+3\vec{\theta}\cdot\vec{B})v^2_{r}-\theta {\cal E}_r{v}_{r}
+2(\theta_{r}B_{r}v^2_{r})\right],
\\
f(v_r)&=& 1-2\vec{\theta}\cdot\vec{B}-3\theta {\cal E}_r v_r,
\\
\Lambda(v_r)&=& 1+\vec{\theta}\cdot\vec{B}-\theta {\cal E}_r v_r,
\\
\Gamma(v_r)&=&1+4\vec{\theta}\cdot\vec{B}-2\theta {\cal E}_r {v}_r,
\\
\Sigma(v_r)&=&\left[\theta {\cal E}_r -({B}_r{v}_r)\theta_{r}-({\theta}_r {v}_r)B_{r}\right]v_{r},
\end{eqnarray}
being $ \theta {\cal E}_r=\theta(\vec{n}\times\vec{E})_{r} $. 
Now, by applying the relation $v_{r}=c_{s}\frac{r^2_{h}}{r^2}$, where $ r_h $ is the radius of the event horizon and making $ c_s=1 $ and so, the metric function of the noncommutative canonical acoustic black hole becomes
\begin{eqnarray}
\label{Fr}
\tilde{\cal F}(r)=\left[ 1-3\vec{\theta}\cdot\vec{B} -(1+3\vec{\theta}\cdot\vec{B}-2\theta_{r}B_{r})
\frac{r^4_{h}}{r^4}
-\theta {\cal E}_r\frac{r^2_{h}}{r^2}\right]
\left[1-2\vec{\theta}\cdot\vec{B}
-3\theta {\cal E}_r\frac{r^2_{h}}{r^2}\right]^{-1/2}.
\end{eqnarray}
Next, we will do our analysis considering the pure magnetic sector first and then we will investigate the pure electric sector.

Hence, for $\theta_r=0$, $\vec{\theta}\cdot\vec{B}=\theta_3B_3\neq 0$,  $\theta {\cal E}_r=0$ (or  $E=0$)  with small $\theta_3B_3$,
\begin{equation} 
\label{Ttb}
T_{H}=\frac{(1+3\theta_3B_3)}
{\sqrt{1-2\theta_3B_3}}\frac{1}{(\pi r_{h})}
=\frac{\left(1+4\theta_3B_3\right)}{(\pi r_{h})}.
\end{equation}
For $\theta=0$ the usual result is obtained. 
Here the temperature has its value increased when we vary the parameter $\theta$.

However, for the temperature in (\ref{Ttb}) we can find the entropy given by
\begin{eqnarray}
S=\int \frac{dE}{T}=\int \frac{dA}{4\pi r_h T_H}=\frac{(1-4\theta_3 B_3)}{4}A,
\end{eqnarray}
where $ A=4\pi r^2_h $ is the horizon area of the canonical acoustic black hole.

At this point, we will consider the situation where $B=0$ and $\theta {\cal E}_r\neq 0$.
So, from (\ref{Fr}), we have
\begin{eqnarray}
\label{FrE}
\tilde{\cal F}(r)=\left[ 1-\frac{r^4_{h}}{r^4}
-\theta {\cal E}_r\frac{r^2_{h}}{r^2}\right]
\left[1-3\theta {\cal E}_r\frac{r^2_{h}}{r^2}\right]^{-1/2}.
\end{eqnarray}
{For this metric the event horizon is obtained by solving the equation below
\begin{eqnarray}
1-\frac{r^4_{h}}{r^4}-\theta {\cal E}_r\frac{r^2_{h}}{r^2}=0,
\end{eqnarray}
or
\begin{eqnarray}
r^4 - \theta {\cal E}_r{r^2_{h}}\,r^2 - r^4_h=0.
\end{eqnarray}
So, solving the above equation, we obtain
\begin{eqnarray}
r_{+}=\left(1+\frac{\theta\mathcal{E}_r}{4}\right)r_h.
\end{eqnarray}
For the Hawking temperature, we find
\begin{eqnarray}
\label{Theb} 
T_{H}&=&\frac{\left[1-\theta {\cal E}_r/2\right]}
{\sqrt{1-3\theta {\cal E}_r}}
\frac{1}{\pi r_{+}}=\frac{\left(1+\theta {\cal E}_r  \right)}{\pi r_{+}},
\\
&=&\frac{\left(1+3\theta {\cal E}_r/4  \right)}{\pi r_{h}}.
\end{eqnarray}
We also noticed that the temperature is increased when we vary the $\theta$ parameter.

For entropy we have }
\begin{eqnarray}
S=\left(1-\theta {\cal E}_r\right)\frac{A}{4},
\end{eqnarray}
where $ A=4\pi r_+ $.

\section{Quantum-corrected Hawking temperature and entropy}
\label{QCTHE}
In this section, we implement quantum corrections in the Hawking temperature and entropy calculation arising from the generalized uncertainty principle and modified dispersion relations.

\subsection{Result using GUP}
At this point, we introduce quantum corrections via the generalized uncertainty principle (GUP) to determine the Hawking temperature and entropy of the canonical acoustic black hole in the Lorentz-violating and noncommutative background.
So, we will adopt the following GUP~\cite{Das:2008kaa,Das:2009hs,Ali:2011fa,Ali:2009zq,Casadio:2014pia,Kempf:1994su,Garay:1994en,Amelino-Camelia:2000cpa,Scardigli:1999jh,Scardigli:2003kr,Scardigli:2014qka,Scardigli:2016pjs}
\begin{eqnarray}
\label{gup}
\Delta x\Delta p\geq \frac{\hbar}{2}\left( 1-\frac{\lambda l_p}{\hbar} \Delta p +\frac{\lambda^2 l^2_p}{\hbar^2} (\Delta p)^2 \right),
\end{eqnarray}
where $\alpha$ is a dimensionless positive parameter and $ l_p $ is the Planck length.

In sequence, without loss of generality, we will adopt the natural units $ G=c=k_B=\hbar=l_p=1 $ and by
assuming that $ \Delta p\sim E $ 
and following the steps performed in~\cite{Anacleto:2019rfn} we can obtain the following relation for the corrected energy of the black hole 
 \begin{eqnarray}
E_{gup}\geq E\left[1-\frac{\lambda}{2(\Delta x)}+ \frac{\lambda^2}{2(\Delta x)^2}+\cdots    \right].
\end{eqnarray}
Thus, applying the tunneling formalism using the Hamilton-Jacobi method, we have the following result for the probability of tunneling with corrected energy $ E_ {gup} $ given by
\begin{eqnarray}
\Gamma\simeq \exp[-2{\rm Im} ({\cal I})]=\exp\left[\frac{-4{\pi}E_{gup}}{\kappa}\right],
\end{eqnarray}
{where $ \kappa $ is the surface gravity}. 
Comparing with the Boltzmann factor, $ e^{-E/T} $, we obtain the following result for the Hawking temperature with quantum corrections 
\begin{eqnarray}
T\leq{T}_H\left[ 1-\frac{\lambda}{2(\Delta x)}+ \frac{\lambda^2}{2(\Delta x)^2}+\cdots   \right]^{-1}.
\end{eqnarray}
So, by applying it to temperature (\ref{temph}), we have the following result
\begin{eqnarray}
\label{Th2}
T=\frac{c^2_s}{\pi\left[\displaystyle r_h-\frac{\lambda}{4}+ \frac{\lambda^2}{8r_h}+\cdots   \right]}.
\end{eqnarray}
Therefore, when $r_h=0$ the singularity is removed and the temperature is now zero.
Next, we analyze the effect of GUP in the Lorentz-violating and noncommutative cases. 

For this case we can calculate the entropy which is given by
\begin{eqnarray}
S=\frac{A}{4c^2_s} - \frac{4\sqrt{\pi}\lambda\sqrt{A}}{4c^2_s}+\frac{\pi\lambda^2\ln A}{8c^2_s}+\cdots.
\end{eqnarray}
So due to the GUP we get a logarithmic correction term for the entropy.

\subsubsection{Violation-Lorentz Case}
In the situation where $ \beta\neq 0 $ and $ \alpha=0 $, the corrected temperature due to GUP is
\begin{eqnarray}
T={T}_H\left[ 1-\frac{\lambda}{4 r_h}+ \frac{\lambda^2}{8r^2_h}+\cdots   \right]^{-1}.
\end{eqnarray}
where
\begin{eqnarray}
T_H=\frac{(1-3\beta)}{\pi r_{h}}.
\end{eqnarray}
Thus, we have
\begin{eqnarray}
\label{Th3}
T=\frac{(1-3\beta)}{\pi\left[\displaystyle r_h-\frac{\lambda}{4}+ \frac{\lambda^2}{8r_h}+\cdots   \right]}.
\end{eqnarray}
Note that when $ r_h\rightarrow 0 $ the Hawking temperature tends to zero, $ T\rightarrow 0 $. 
In the absence of the GUP the temperature, $ T_H $, diverges when $ r_h=0 $. Therefore, we observe that the GUP has the effect of removing the singularity at $ r_h=0 $ in the Hawking temperature of the acoustic black hole.

Now computing the entropy, we find
\begin{eqnarray}
S=\left(1+3\beta\right)\left[\frac{A}{4} - \frac{4\sqrt{\pi}\lambda\sqrt{A}}{4}
+\frac{\pi\lambda^2\ln A}{8}+\cdots\right].
\end{eqnarray}

\begin{figure}
\includegraphics[scale=0.5]{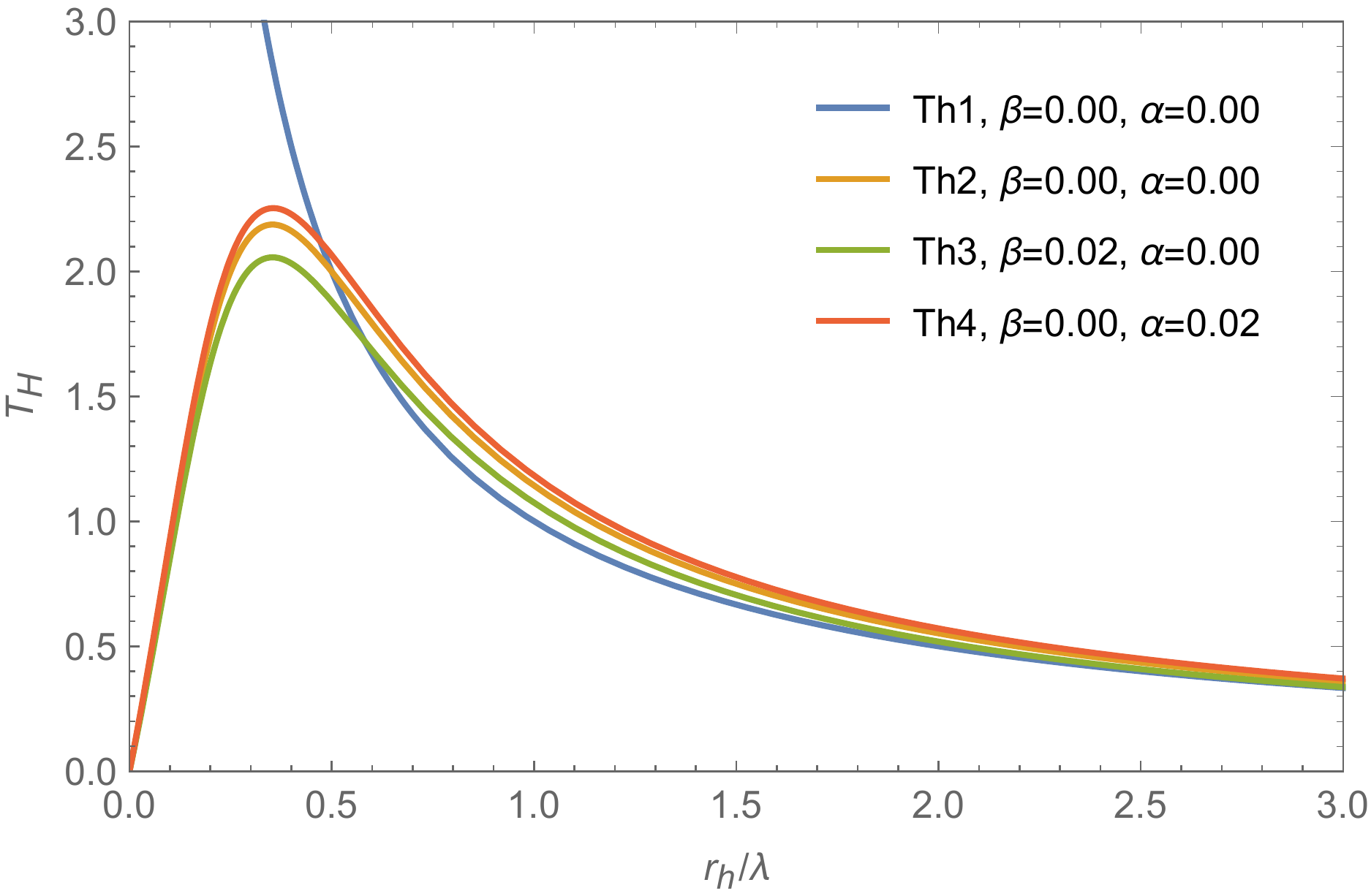}
\caption{The Hawking temperature $ T_H \rightarrow \pi\lambda T_H$ in function of $ r_h/\lambda $.
Note that the temperature $ Th1 $ (\ref{temph}) diverges to $ r_h/\lambda\rightarrow 0 $  and the temperatures $ Th2 $ (\ref{Th2}) , $ Th3 $ (\ref{Th3}) and $ Th4 $ (\ref{Th4}) reach maximum values, and then decreases to zero as
$ r_h/\lambda\rightarrow 0 $.}
\label{thgup}
\end{figure}

For $ \beta=0 $ and $ \alpha\neq 0 $, we have
\begin{eqnarray}
T=\frac{(1+3\alpha/2)}{\pi\left[\displaystyle \tilde{r}_{+}-\frac{\lambda}{4}+ \frac{\lambda^2}{8\tilde{r}_{+}}+\cdots   \right]}.
\end{eqnarray}
In terms of $ r_h $, we obtain 
\begin{eqnarray}
\label{Th4}
T=\frac{(1+3\alpha/2)}{\pi\left[\displaystyle {r}_{h}\left(1-\frac{\alpha}{4}\right)-\frac{\lambda}{4}
+ \frac{\lambda^2}{8{r}_{h}}\left(1+\frac{\alpha}{4}\right)+\cdots   \right]}.
\end{eqnarray}
In this situation, we can also verify the effect of the GUP on the temperature that goes to zero 
when $ r_h\rightarrow 0 $  $ (\tilde{r}_+\rightarrow 0) $. 
In addition, we note that in both cases the Hawking temperature reaches a maximum value before going to zero, 
{ as we can see in Fig.~\ref{thgup}.} Therefore, presenting a behavior analogous to what happens with the corrected Hawking temperature of the Schwarzschild black hole.

For entropy, we obtain
\begin{eqnarray}
S=\left(1-\frac{3\alpha}{2}\right)\left[\left(1-\frac{\alpha}{4}\right)\frac{A}{4} - \frac{4\sqrt{\pi}\lambda\sqrt{A}}{4}
+\left(1+\frac{\alpha}{4}\right)\frac{\pi\lambda^2\ln A}{8}+\cdots\right].
\end{eqnarray}
Again we find a logarithmic correction term and also the contribution of the $\alpha$ parameter to the entropy.

\subsubsection{ Noncommutative Case}
For the magnetic sector, the GUP-corrected Hawking temperature is given by
\begin{eqnarray}
T=\frac{(1+4\theta_3 B_3)}{\pi\left[\displaystyle r_h-\frac{\lambda}{4}+ \frac{\lambda^2}{8r_h}+\cdots   \right]}.
\end{eqnarray}
Note that, the GUP acts as a temperature regulator by removing the singularity when $r_h=0$. In addition, the temperature goes through a maximum value point before going to zero for $r_h=0$.

In this case entropy is given by
\begin{eqnarray}
S=\left(1-4\theta_3 B_3\right)\left[\frac{A}{4} - \frac{4\sqrt{\pi}\lambda\sqrt{A}}{4}
+\frac{\pi\lambda^2\ln A}{8}+\cdots\right].
\end{eqnarray}

Next, for the electrical sector, we find the following GUP-corrected Hawking temperature
\begin{eqnarray}
T=\frac{(1+\theta \mathcal{E}_r)}{\pi\left[\displaystyle r_{+}-\frac{\lambda}{4}+ \frac{\lambda^2}{8r_{+}}+\cdots   \right]}.
\end{eqnarray}
In terms of $r_h$, the temperature becomes
\begin{eqnarray}
T=\frac{(1+\theta \mathcal{E}_r)}{\pi\left[\displaystyle r_{h}\left(1+\frac{\theta \mathcal{E}_r}{4}\right) -\frac{\lambda}{4}+ \frac{\lambda^2}{8r_{h}}\left(1-\frac{\theta \mathcal{E}_r}{4}\right)+\cdots   \right]}.
\end{eqnarray}
Hence, as has been verified in the violating-Lorentz case, here in both cases the temperature-corrected magnetic and electric sectors have the singularity removed when the horizon radius goes to zero. 
Also, in this case we can observe that the temperature reaches a maximum value and then goes to zero when the horizon radius is zero.

At this point, when determining the entropy, we have
\begin{eqnarray}
S=\left(1-\theta \mathcal{E}_r\right)\left[\left(1+\frac{\theta \mathcal{E}_r}{4}\right)\frac{A}{4} - \frac{4\sqrt{\pi}\lambda\sqrt{A}}{4}
+\left(1-\frac{\theta \mathcal{E}_r}{4}\right)\frac{\pi\lambda^2\ln A}{8}+\cdots\right].
\end{eqnarray}

\subsection{Result using modified dispersion relation}
\begin{figure}
\includegraphics[scale=0.5]{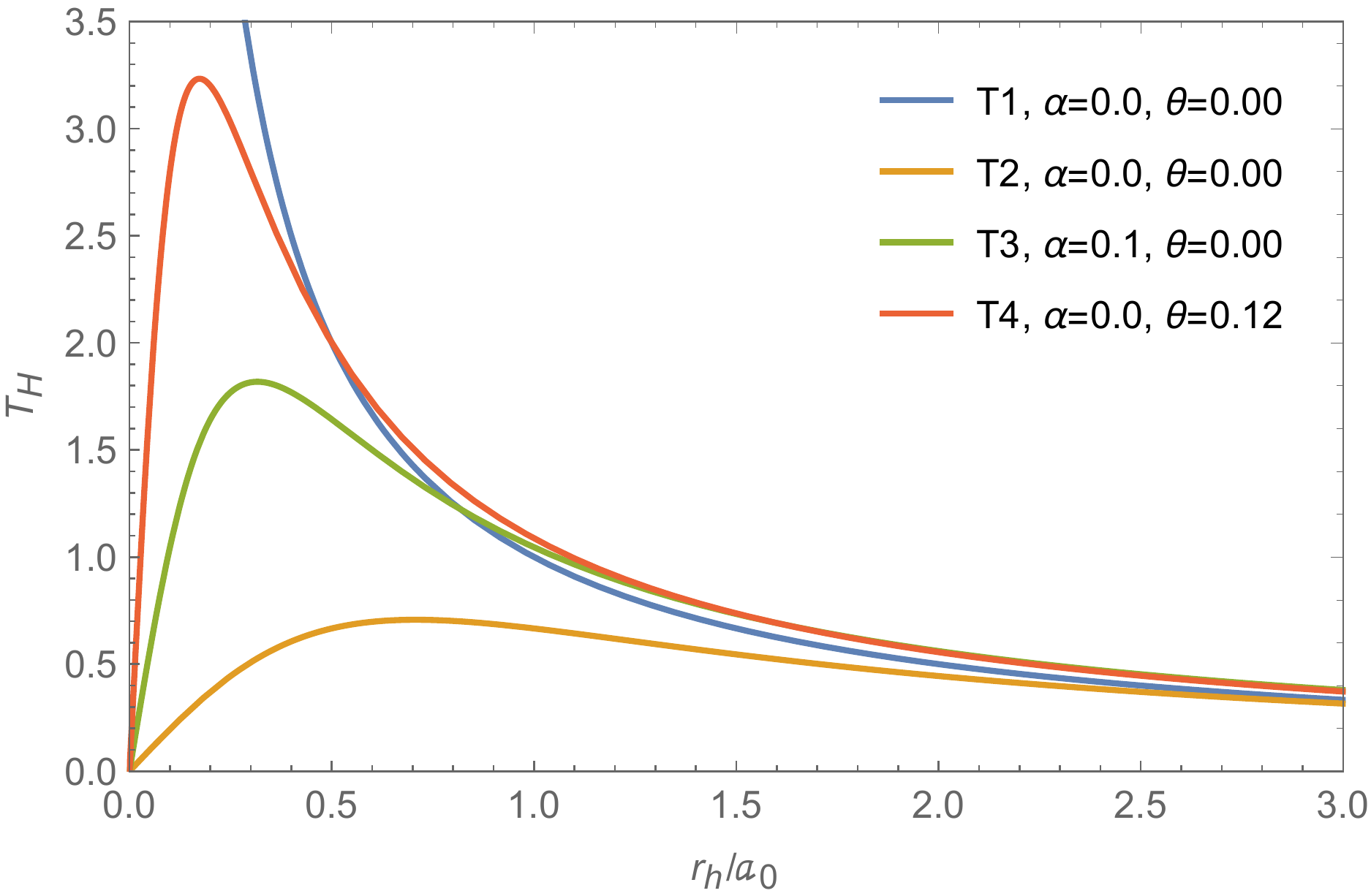}
\caption{The Hawking temperature $ T_H \rightarrow \pi a_0 T_H$ in function of $ r_h/a_0 $.
Note that the temperature $ T1 $ (\ref{temph}) diverges to $ r_h/a_0\rightarrow 0 $  and the temperatures $ T2 $ (\ref{T2}) , $ T3 $ (\ref{T3}) and $ T4 $ (\ref{T4}) reach maximum values, and then decreases to zero as $ r_h/a_0\rightarrow 0 $.}
\label{thdr}
\end{figure}

{Near the event horizon the dispersion relation (\ref{rdisp})  becomes
\begin{eqnarray}
\label{rldh}
\omega= E\left(1+\frac{a^2_0}{2r^2_h} \right),
\end{eqnarray}
where $ a_0 $ is a parameter with length dimension.
By assuming $ k\sim \Delta k\sim 1/\Delta x=1/r_h $, we can write
\begin{eqnarray}
\omega= E\left(1+\frac{a^2_0\,k^2}{2} \right).
\end{eqnarray}
Thus, in terms of the energy difference, we have
\begin{eqnarray}
\frac{\Delta E}{E}=\frac{\omega - E}{E}=\frac{a^2_0\,k^2}{2}.
\end{eqnarray}

Next, by using the Rayleigh's formula that relates the phase and group velocities
\begin{eqnarray}
v_g=v_p + k \frac{dv_p}{dk},
\end{eqnarray}
where the phase velocity ($v_p$) and the group velocity ($v_g$) are given by
\begin{eqnarray}
v_p=\frac{{\omega}}{k}=1+\frac{a^2_0\,k^2}{2},
\end{eqnarray}
and 
\begin{eqnarray}
v_g=\frac{d{\omega}}{dk}=1+\frac{3a^2_0\,k^2}{2}.
\end{eqnarray}
However, we find an expression for the velocity difference as following
\begin{eqnarray}
\frac{v_g - v_p}{v_p}={a^2_0\,k^2},
\end{eqnarray}
which corresponds to the supersonic case ($ v_g > v_p $).

Furthermore, the Hawking temperature (\ref{temph}) can be corrected by applying the dispersion ratio (\ref{rldh}), i.e.
\begin{eqnarray}
\label{T2}
T_H=\frac{c^2_s}{\displaystyle\pi\left(r_n + \frac{a^2_0}{2r_h}  \right)}.
\end{eqnarray}
Note that, the singularity is removed when $ r_h=0 $ and the temperature vanishes. }
In addition, the temperature reaches a maximum value before going to zero.
as we can see in Fig.~\ref{thdr}.

Now, by calculating the entropy, we find
\begin{eqnarray}
S=\frac{A}{4c^2_s}+\frac{2\pi a_0^2\ln A}{4c^2_s}.
\end{eqnarray}
Here a logarithmic correction term arises in entropy on account of the modified dispersion relation.

In order to correct the Hawking temperature and entropy for the Lorentz-violating and non-commutative cases, we will apply the modified dispersion relations obtained in Refs.~\cite{Anacleto:2010cr,Anacleto:2011bv}.

\subsubsection{Violation-Lorentz Case}
In the situation where $\beta=0$ and $\alpha\neq 0$, we have the following dispersion relation:
\begin{eqnarray}
\omega=E\left(1+\frac{\alpha}{2}+\frac{\alpha a_0^2}{\tilde{r}^2_{+}} \right).
\end{eqnarray}
So for temperature (\ref{tha}), we get
\begin{eqnarray}
\label{T3}
T=\frac{(1+3\alpha/2)}{\pi\left[\displaystyle \tilde{r}_{+}+\frac{\alpha}{2}+ \frac{\alpha a_0^2}{\tilde{r}_{+}} \right]}.
\end{eqnarray}
Furthermore, the result shows that the temperature reaches a maximum point and then goes to zero when the horizon radius is zero.
Moreover, entropy is given by
\begin{eqnarray}
S=\left(1-\frac{3\alpha}{2}\right)\left[\left(1-\frac{\alpha}{4}\right)\frac{A}{4} + \frac{2\sqrt{\pi}\alpha\sqrt{A}}{4}
+\frac{4\pi\alpha a^2_0\ln A}{4}\right].
\end{eqnarray}
Again due to the contribution of the modified dispersion relation, a logarithmic correction term arises in the entropy.

\subsubsection{ Noncommutative Case}
At this point we consider the dispersion ratio for the pure electrical sector. So we have
\begin{eqnarray}
\omega=E\left(1+\frac{\theta\mathcal{E}_1 a_0^2}{4{r}^2_{+}} \right).
\end{eqnarray}
For the temperature (\ref{Theb}), we find
\begin{eqnarray}
T=\frac{\left(1+\theta {\cal E}_r  \right)}{\pi\displaystyle\left(r_{+} + \frac{\theta\mathcal{E}_1 a_0^2}{4{r}_{+}} \right)},
\end{eqnarray}
which in terms of $r_h  $ becomes
\begin{eqnarray}
\label{T4}
T=\frac{\left(1+\theta {\cal E}_r  \right)}{\pi\displaystyle\left[\left(1+\frac{\theta \mathcal{E}_r}{4}\right)r_{h} + \frac{\theta\mathcal{E}_1 a_0^2}{4{r}_{h}} \right]}.
\end{eqnarray}
Here, we can see that the temperature goes through a maximum value before going to zero for $ r_h=0 $.

Hence, the result for entropy is
\begin{eqnarray}
S=\left(1-\theta \mathcal{E}_r\right)\left[\left(1+\frac{\theta \mathcal{E}_r}{4}\right)\frac{A}{4}
+\frac{\pi\theta\mathcal{E}_1 a_0^2\ln A}{4}\right].
\end{eqnarray}
In the above equation a logarithmic correction term arises in entropy as a consequence of the noncommutativity effect on the dispersion relation.

\section{conclusions}
\label{conclu}

In summary, in this work, we have reviewed the steps to generate relativistic acoustic metrics in the Lorentz-violating 
and noncommutative background.
In particular, we have considered the modified canonical acoustic metric due to the contribution of terms violating Lorentz symmetry and noncommutativity; to examine Hawking radiation and entropy. 
Moreover, we have verified, in the calculation of the Hawking temperature, that due to the presence of the GUP and the modified dispersion relation, the singularity is removed. In addition, we have shown that in these cases, the temperature reaches a maximum value and then vanishes when the horizon radius goes to zero. 
Furthermore, entropy has been computed, and we show that logarithmic correction terms are generated due to the GUP and also the modified dispersion relation. 
Therefore, the presented results show a behavior similar to what happens in the case of the Schwarzschild black hole.

\acknowledgments

We would like to thank CNPq, CAPES and CNPq/PRONEX/FAPESQ-PB (Grant nos. 165/2018 and 015/2019),
for partial financial support. MAA, FAB and EP acknowledge support from CNPq (Grant nos. 306398/2021-4, 312104/2018-9, 304290/2020-3).

\end{document}